\documentclass[11pt,twoside]{article} % Specifies the document style.
\pdfoutput=1
\usepackage{times,fancyhdr}
\usepackage{amsmath}
\usepackage{amssymb}
\usepackage{color}
\usepackage{graphics}
\usepackage{graphicx}
\date{}

\title{Atmospheric muons: experimental aspects}

\author{S. Cecchini (1) and M. Spurio (1,2) \\
(1) INFN, sezione di Bologna \\
(2) Dipartimento di Fisica dell'Universit\`a di Bologna}
\begin{document} % End of preamble and beginning of text.

\pagestyle{fancy}
\fancyhead{} % clear all header fields
\fancyhead[EC]{ S. Cecchini and M. Spurio}
\fancyhead[EL,OR]{\thepage}
\fancyhead[OC]{Atmospheric muons: experimental aspects }
\fancyfoot{} % clear all footer fields
\renewcommand\headrulewidth{0.5pt}
\addtolength{\headheight}{2pt} % make space for the rule

\maketitle % Produces the title.

\baselineskip=11.6pt

\begin{abstract}
We present a review of atmospheric muon flux and energy spectrum measurements over almost six decades of muon momentum. Sea-level and underground/water/ice experiments are considered. 
 Possible sources of systematic errors in the measurements are examinated. The characteristics of underground/water muons (muons in bundle, lateral distribution, energy spectrum) are discussed. The connection between the atmospheric muon and neutrino measurements are also reported. 
\end{abstract}
\newpage
%\baselineskip=14pt

%%%%%%%%%%%%%%%%%%%%%%%%%%%%%%%%%%%%%%%%%%%%%%%%%%%%%%%%%%%%%%%%%%%%%%%%%
%%%%%%%%%%%%%%%%%%%%%%%%%%%%%%%%%%%%%%%%%%%%%%%%%%%%%%%%%%%%%%%%%%%%%%%%%
%%%%%%%%%%%%%%%%%%%%%%%%%%%%%%%%%%%%%%%%%%%%%%%%%%%%%%%%%%%%%%%%%%%%%%%%%
%%\introduction  
%% \introduction[modified heading if necessary]
%%%''''''''
% {}``naive''
\section{Introduction}

Primary cosmic rays (CRs) are particles accelerated at astrophysical sources continuously bombarding the Earth. When entering the Earth's atmosphere, primary CRs interact with the air nuclei and produce 
fluxes of secondary, tertiary (and so on) particles. All these particles
together create a cascade, called air shower. 
As the cascade develops longitudinally the particles are less and less energetic since the energy of the incoming CR is split and redistributed among more and more participants. The transverse momenta acquired by the secondaries cause the particles to spread laterally as they propagate in the target. 
 Muons predominantly originate from the decay of secondary charged pions and kaons. The most important decay channels and respective branching ratios are:
\begin{subequations}
\label{eq:meson}
\begin{equation}
\pi^\pm \rightarrow \mu^\pm \overset{(-)~}{\nu_\mu} \quad (\sim 100\%) 
\end{equation}
\begin{equation}
K^\pm \rightarrow \mu^\pm \overset{(-)~}{\nu_\mu} \quad (\sim 63.5\%) 
\end{equation}
\end{subequations}

Atmospheric muons are the most abundant charged particles arriving at sea level and the only ones able to penetrate deeply underground.
The reason relies on their small energy loss in the whole atmosphere ($\sim$ 2 GeV), the relatively long lifetime and the fairly small interaction cross section. Because of the close relation between muon and neutrino production, the parameters characterizing muon physics can provide important information on atmospheric neutrino flux.

An important parameter to describe the interactions and the subsequent propagation of the particles produced is the atmospheric depth $X$, measured in g/cm$^2$, defined as the integral in altitude of the atmospheric density above the observation level $h$:
\begin{equation}
X= \int_h^\infty \rho(h^\prime)dh^\prime \simeq X_0 e^{-h/h_0}
\label{eq:atmh}
\end{equation}
In the last step, an approximation for an isothermal atmosphere was used, where $X_0 = 1030$ g/cm$^2$ is the atmospheric depth at sea level and $h_0\simeq 8.4$ km is the scale height (for mid latitudes) in the atmosphere. Eq. (\ref{eq:atmh}) is valid for vertically incident particles. For zenith
angles $< 60^\circ$, for which the Earth surface can be approximated as flat, the atmospheric depth is scaled with $1/\cos\theta$, giving the \textit{slant depth}. For larger zenith angles, the curvature of the Earth has to be accounted for. The atmospheric profile gives a total horizontal
atmospheric depth of about 36000 g/cm$^2$.

The air shower is described by a set of coupled cascade equations with boundary conditions at the top of the atmosphere to match the primary spectrum. 
Using the transport equations, analytic expressions of the cascade can be constructed. The solutions of these equations allow to compute the differential particle flux anywhere within the atmospheric target. Some approximate analytic solutions are valid in the limit of high energies \cite{gai90},\cite{gai02},\cite{lipa93}. 
Numerical or Monte Carlo calculations are needed to account accurately for decay and energy loss processes, and for the energy-dependences of the cross sections and of the primary spectral index. 
The nucleon mean free path $\lambda_N$ in atmosphere is given (in  g/cm$^2$ units) by 
\begin{equation}
\label{eq:lambda}
\lambda_N= {A m_p \over \sigma_N^{air}}
\end{equation}
where $\sigma_N^{air}$ is the interaction cross section of nucleon in air, $A$ is the mean mass number of air nuclei and $m_p$ the proton mass. For nucleons in the TeV range, $\sigma_N^{air} \simeq 300$ mb. In the context of air shower development, the energy-dependent cross section for an inelastic collision of a nucleon with an air nucleus is assumed to be constant. The atmosphere of the Earth consists mainly of nitrogen and oxygen: the interaction target for the primary beam is half protons and half neutrons. 
Assuming an average atmospheric nucleus with $A\sim 14.5$,  $\lambda_N\simeq 80$ g/cm$^2$. The total vertical atmospheric depth is about 1000 g/cm$^2$ and it corresponds to more than 11 interaction lengths.

Most muons are produced through processes (\ref{eq:meson}) high in the atmosphere in the first few interaction lengths. 
The decay mean free path of pions $d_\pi$ in units of slant depth, is defined as:
\begin{equation}
\label{eq:dpi}
{1 \over d_\pi} = { m_\pi c^2 h_0 \over E c \tau_\pi X \cos\theta }
= { \epsilon_\pi \over E X \cos\theta }
\end{equation}
where E, $m_\pi,\tau_\pi$ are the pion energy, mass and lifetime, respectively. A similar relation holds for the kaon. Decay or interaction dominates depending on whether $1/d_\pi$ or $1/\lambda_\pi$ is larger. $\lambda_\pi $ is defined through Eq. (\ref{eq:lambda}) replacing $\sigma_N$ with $\sigma_\pi$. 
At the critical energy $E=\epsilon_\pi = m_\pi c^2 h_0/\tau_\pi c=115$ GeV the interaction probability in the atmosphere equals the decay probability. 

%%%%%%%%%%
As for the pion, all long-lived unstable particles \cite{bgs} 
%($10^{-8}\le \tau \le 10^{-10}$ s) 
are subject to competition between interaction and decay as they propagate in the atmosphere. The probability for either process to occur depends on the lifetime of the particle and is a function of its kinetic energy and on the local atmospheric density, which is a function of altitude.
This interrelationship is responsible for the zenith angle enhancement of the bulk of the muons in air showers. 

The zenith angle enhancement phenomenon does not affect the distribution of muons produced in semileptonic decays of charmed mesons, like $D^\pm, D^0$ and others. As the lifetime of charmed particles is smaller than $\sim 10^{-12}$ s (\textit{prompt decays}), they yield so-called \textit{prompt} (or \textit{direct}) muons that are in general highly energetic for kinematic reasons. Since the production cross section of charmed mesons in proton-nucleon interactions is rather small, $D$ decays contribute significantly only at very high energies.
%%%%%%%%%

%%%%%%%%%%%%%%%%%%%%%%%%%%%%%%%%%%%%%%%%%
\section{Cosmic rays at the sea level}
\begin{figure}[t]
\vspace*{2mm}
\begin{center}
\includegraphics[width=9.3cm]{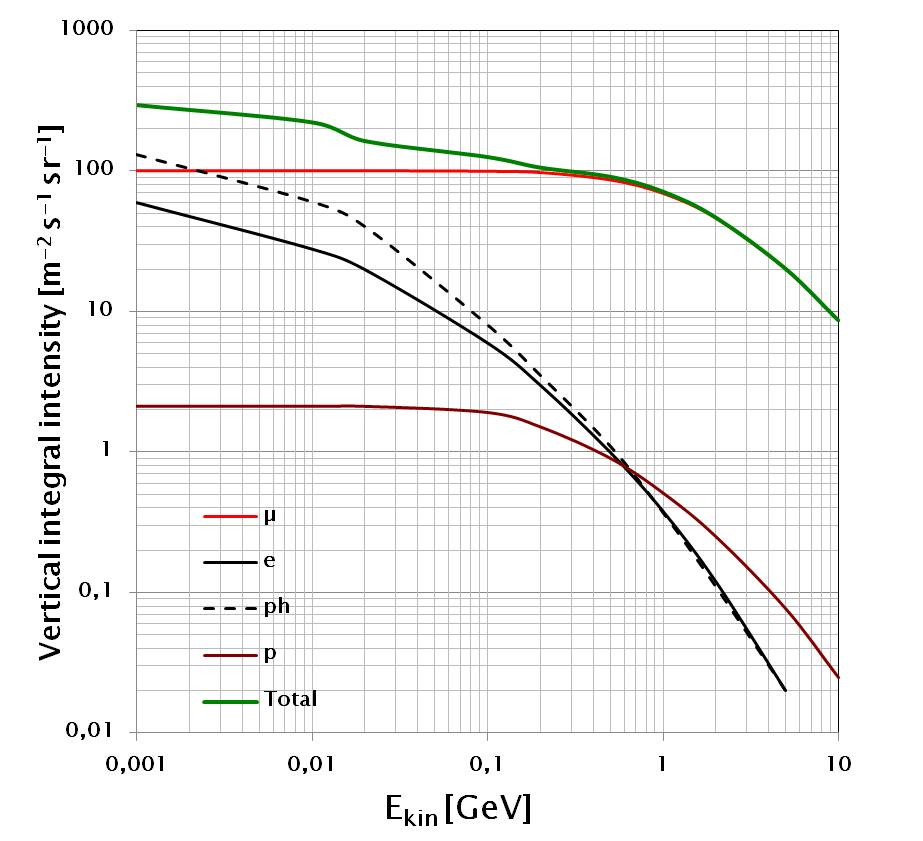}
\end{center}
\caption{\label{fig:ce-ek}
Integral fluxes averaged over the 11-year solar cycle of $\mu, e, p $ and photons ($ph$) arriving at geomagnetic latitudes $\sim 40^\circ$ vs. their kinetic energy.}
\end{figure}

%%%%%%%%%%%%%%%%%%%%%%%%%%%%%%%%%%%%%%%%%
Muons are the dominant component of charged particles at sea level. The integral fluxes of particles arriving at geomagnetic latitudes $\sim 40^\circ$ vs. their kinetic energy are presented in Fig. \ref{fig:ce-ek}. Fluxes are averaged over the 11-year solar cycles. 
The muon flux with $E_\mu> 1$ GeV through a horizontal area amounts to roughly one particle per cm$^2$ and per minute:
$I_v(E_\mu > 1 \textrm{ GeV})\sim 70$ m$^{-2}$ s$^{-1}$ sr$^{-1}$ \cite{gri10}.
%%%%%%%%%%%%%%%%%%%%%%%%%
%%%%%%%%%%%%%%%%%%%%%%%%%

The muon energy spectrum at sea level is a direct consequence of the meson source spectrum. 
Below the few GeV energy range, the muon decay probability cannot be neglected. A muon of 1 GeV has a Lorentz factor $\gamma = E_\mu/m_\mu c^2\sim 10$ and it has a mean decay length 
$d_\mu = \gamma \tau_\mu c \sim 6$ km. Since pions are typically produced at altitudes of 15 km and decay relatively fast (for $\gamma = 10$ the decay length is $d_\pi\sim 78$ m, which is almost the same value of $\lambda_\pi$), the daughter  muons do not reach the sea level but rather decay themselves or get absorbed in the atmosphere.

At higher energies, the situation changes. For pions of 100 GeV ($d_\pi\sim 5.6$ km, corresponding to a column density of 160 g/cm$^2$ measured from the production altitude) the interaction probability starts to dominate over decay. Pions of these energies will therefore produce further tertiary pions in subsequent interactions, which will also decay eventually into muons, but providing muons of lower energy. Therefore, the muon spectrum at high energies is always steeper compared to the parent pion spectrum.

The muon energy and angular distribution is the effect of a convolution of production spectrum, energy losses in the atmosphere and decay.
The competition of decay and interaction plays a crucial role and the relative importance of the two processes depends on energy. The mean energy of muons at the ground is about 4 GeV \cite{rpp}. 

Three different energy regions in the muon spectrum are distinguishable:

\vskip 0.2cm
\noindent $\bullet$ $ E_\mu  \le \epsilon_\mu$, where $\epsilon_\mu \sim 1$ GeV. In this case, muon decay and muon energy loss are important and must be taken into account. Semi-analytical solutions (as those used in the higher energy range) overestimate the flux. The energy spectrum is almost flat, starting to steepen gradually in the same way as the primary spectrum above 10 GeV.

\vskip 0.2cm
\noindent $\bullet$ $\epsilon_\mu \le E_\mu \le \epsilon_{\pi,K}$, where $\epsilon_{\pi}$= 115 GeV and $\epsilon_{K}$ = 850 GeV are the critical energies for the vertical directions. In this energy range, almost all mesons decay and the muon flux has the same power law of the parent mesons, and hence of the primary CRs. The muon flux is almost independent on the zenith angle.

\vskip 0.2cm
\noindent $\bullet$ $E_\mu \gg \epsilon_{\pi,K}$. The meson production spectrum has the same power law dependence of the primary CRs, but the rate of their decay steepen one power of $E_\mu$ since the pion and kaon decay probability is suppressed. 
The thickness of the atmosphere is not large enough for pions to decay, since the high Lorentz factor. For $E > \epsilon_{\pi}$  the inclined muon spectrum is  flatter than the vertical one and the muon  flux is respectively higher.

In the intermediate and high energy region (above 100 GeV) and for zenith angle $\theta < 60^\circ$ an approximate formula holds:
\begin{equation}
\label{eq:gaisser}
{dN_\mu (E_\mu,\theta) \over dE_\mu d\Omega} = \newline
A E_\mu^{-\gamma} \biggl(          { 1 \over 1+({aE_\mu \over \epsilon_\pi }) \cos\theta } + { B \over 1+({ bE_\mu \over \epsilon_K }) \cos\theta } \biggr)
\end{equation}
where the scale factor $A$,  the power index $\gamma$, the balance factor $B$ (which depends on the ratio of muons produced by kaons and pions), and the $a,b$ coefficients are adjustable parameters. Different best estimates of these parameters were published by several authors. For a review, see \cite{geo}.
Because pions decay more easily in non-vertical showers, a zenith angle $\theta$ factor enters in the formula and muons at large angles have a flatter energy spectrum.

%%%%%%%%%%%%%%%%%%%%%
\subsection{Angular distribution at sea level}
%%%%%%%%%%%%%%%%%%%%%
The muon intensity from horizontal directions at low energies is naturally reduced because of muon decays and absorption effects in the thicker atmosphere at large zenith angles. At high energy the parent particles of muons travel relatively long distances in rare parts of the atmosphere. As a consequence, their decay probability is increased compared to the interaction probability.

\begin{figure}[t]
\vspace*{2mm}
\begin{center}
\includegraphics[width=9.3cm]{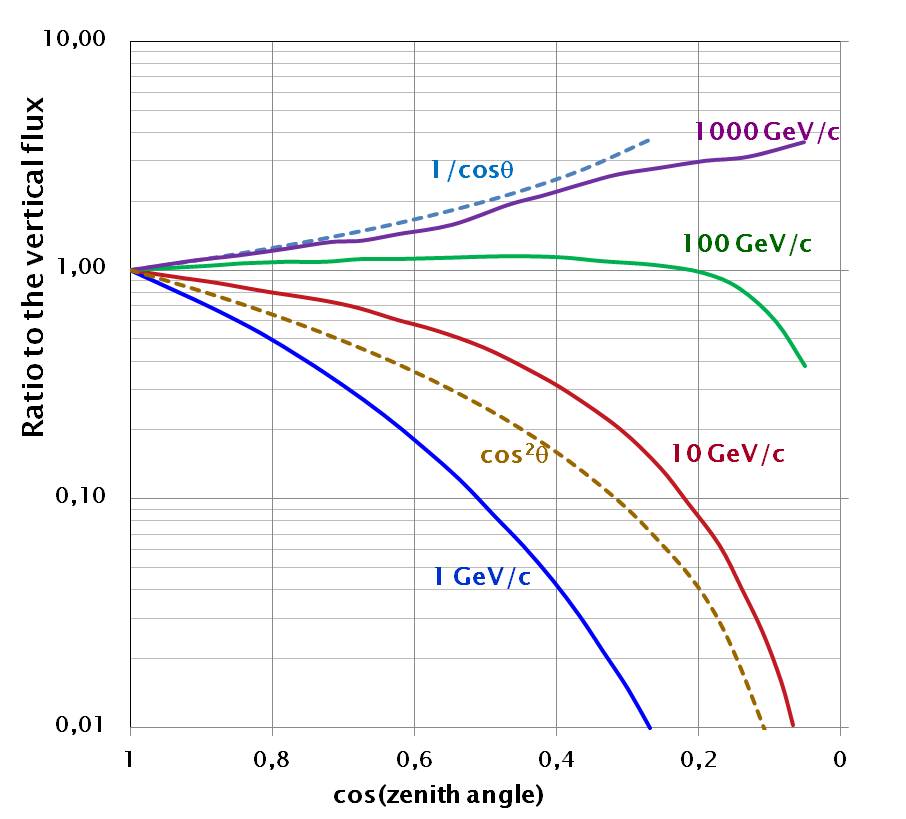}
\end{center}
\caption{\label{fig:zenith}
Angular distribution of muons at the ground for different muon energies. The overall angular distribution of muons measured at sea level is $\propto \cos^2\theta$, which is characteristic of muons with $E_\mu \sim 3$ GeV. 
At lower energy the angular distribution becomes increasingly steep, while at higher energy it  flattens, approaching a $1/\cos\theta$ distribution for $E_\mu \gg \epsilon_\pi$ and $\theta < 60^\circ$. At large angles low energy muons decay before reaching the surface and high energy pions decay before they interact, thus the average muon energy increases. The estimate of the angular distribution is based on a Monte Carlo and accounts for the curvature of the Earth atmosphere.}
\end{figure}

Fig. \ref{fig:zenith} gives a quantitative description of this effect. Muons below the few GeV/c momentum range fade fairly quickly with increasing zenith angle, with dependence $\propto \cos^n\theta$, where $n\sim 2\div 3$. 

 The flux of muons in the 100 GeV/c range is relatively flat up to $\cos\theta \simeq 0.2$ and then quickly declines. At 1 TeV/c the flux monotonically increases with the zenith angle, approaching the $1/\cos\theta$ dependence. The flux of TeV muons is particularly sensitive to large values of the zenith angle. When approaching the horizontal direction, a small difference in $\cos\theta$ changes appreciably the thickness and the density profile of the atmosphere and the corresponding muon energy spectrum. For this reason the measurements of almost horizontal muons is very difficult.
%%%%%%%%%%%%%%%%%%%%%%
%%%%%%%%%%%%%%%%%%%%%%%

%%%%%%%%%%%%%%%%%%%%%
\section{Measurements at sea level}
%%%%%%%%%%%%%%%%%%%%%
Measurements performed at ground level offer the advantage of a high stability, large collecting factor and long exposure time due to relatively favourable experimental conditions. Sea level data offer the possibility to perform a robust check of the reliability of existing Monte-Carlo codes.

Many of the experiments devoted to the measurement of the muon momentum
spectra and intensity (vertical and inclined directions) have been carried out since the '70s. The results are often in disagreement with one another; the discrepancies are significantly larger that the experimental reported errors. Recently new instruments, mainly spectrometers designed for balloon experiments or used primarily in CERN LEP and LHC experiments and used also to CR studies, have added new valuable information.

The vertical muon intensity at sea level is a quantity that varies with the geomagnetic latitude, altitude, solar activity and atmospheric conditions. When comparing muon observations at low energies ($<$ 20 GeV/c) it is important to know the year and location where the measurements were made  \cite{cs}. 

The \textbf{geomagnetic field} tends to prevent low energy cosmic rays from penetrating through the magnetosphere down to the Earth's atmosphere. At any point on the Earth one can define a threshold (or cut-off) rigidity for cosmic rays arriving at a particular zenith and azimuth angle. Primary nuclei having lower rigidity are deflected by the action of the geomagnetic field and do not produce muons at that latitude. The cut-off values range from less than 1 GV near the geomagnetic poles to about 16 GV for vertical particles near the equator. As CR primaries are predominantly protons and nuclei, it results that at a given location the intensity from the West is stronger than that from the East. 
The geomagnetic effects are important for sea level muons up to about $E_\mu\sim  5$ GeV, and the effect is larger at higher altitudes. 

The 11 year \textbf{solar cycle} influences the primary CR spectrum at the top of the atmosphere, as the configuration of the Interplanetary Magnetic Field varies. It results that the cosmic ray flux is significantly modulated up to energies of about 20 GeV.

Most experiments are not exactly performed at sea level. A correction to take into account the dependence of the flux from the \textbf{altitude} must be included. For muon momenta above 10 GeV and altitudes $H$  less than $\sim $ 1000 m the vertical muon flux can be parameterized as
$I_\mu({H}) = I_\mu(0) e^{-H/L(p)}$, where $L(p)= 4900+ 750p$  is a scale factor (in meters) which depends on muon momentum, $p$ (measured in GeV/c).

Changes in \textbf{pressure and temperature} in the atmosphere above the detector produce variations which will be considered in \S \ref{sec:tp}.

%%%%%%%%%%%%%%%%%%%%%
\subsection{Experimental setups\label{sec:setups}}
%%%%%%%%%%%%%%%%%%%%%

Different experimental methods were used to measure the muon flux and energy spectrum. 

Muon telescopes are made of several charged particle detectors arranged along a straight line and interlaid by one or more layers of absorbing material. In some experiments the detector and absorber are in a rigid construction which could be rotated in zenith and azimuth direction, allowing the selection of muons from a given direction of celestial hemisphere. The quantity of material (in g/cm$^2$) travelled by muons in such a telescope is approximately constant and it sets the muon energy threshold.

\begin{figure}[t]
\vspace*{2mm}
\begin{center}
\includegraphics[width=10.0cm]{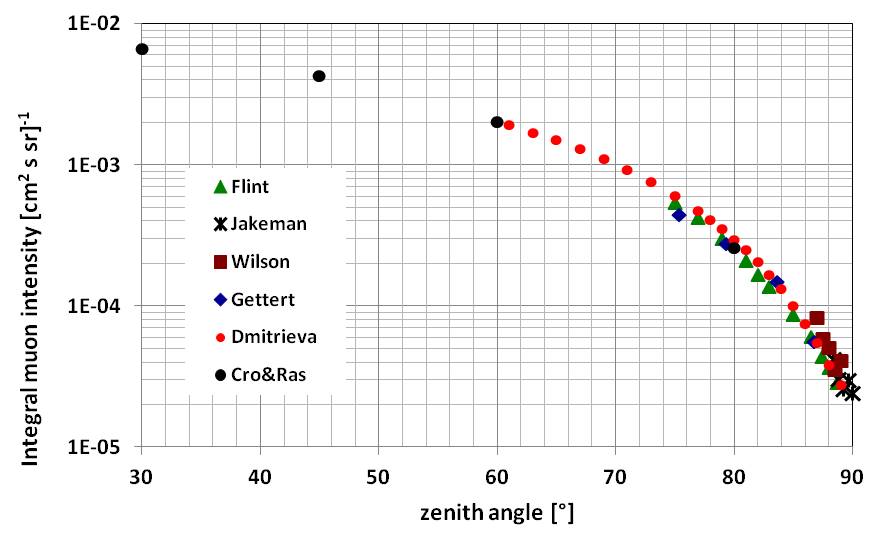}
\end{center}
\caption{\label{fig:zeni}
Variation of absolute integral intensity of muons at sea level with the zenith angle. The momentum threshold $p_T$ is slightly different for each experiment and ranges from 0.1 to 1 GeV/c. The points are normalized assuming $p_T=0.35$ GeV/c. The points correspond (from top to bottom): \cite{flint}, \cite{jake}, \cite{wilson}, \cite{gett}, \cite{dmit2}, \cite{cro}. }
\end{figure}

Multi-directional muon telescopes generally consist of at least two layers of segmented muon detectors. The coincidence of signals between two counters in upper and bottom layers determines the direction of muon arrival. 
The quantity of material crossed by the particle in such detectors increases with increase of zenith angle, so the threshold energy for multidirectional muon telescopes depends on $\theta$. 
A compilation of measurements of the muon flux at latitudes between 52$^\circ$ and 56$^\circ$ as a function of the zenith angle by different experiments is reported in Fig. \ref{fig:zeni}. 
%%%%%%%%%%%%%%%%%%%%%%
%%%%%%%%%%%%%%%%%%%%%%%

%ESPERIMENTI CON MAGNETI (CAPRICE,BESS, L3,…)
The muon energy spectrum has been extensively measured, mainly by solid iron magnet spectrometers. For these detectors, the multiple scattering plays an important role in the momentum resolution, particularly at low energies. 
Recently, measurements have been performed using low mass superconducting magnet spectrometers designed as a balloon-borne apparatus for cosmic ray studies. They represent a sort of second generation experiments. 

The atmospheric muon flux and energy spectrum was also measured using the precise muon spectrometer of the L3 detector which was located at the LEP collider at CERN. This apparatus collected muons $\sim$30 m below a stratified rock overburden, and with dimensions much larger than previous experiments (the volume of the region with a magnetic field of 0.5 T was $\sim$1000 m$^3$).

Finally, information about the muon flux with momenta larger than few TeV  have been extracted from underground measurements, see \S \ref{sec:under}.
Comprehensive review of various types of particle detectors used for cosmic ray studies can be found in \cite{dor04}.

%%%%%%%%%%%%%%%%%%%%%
\subsection{Momentum spectra at the vertical direction\label{sec:muvert}}
%%%%%%%%%%%%%%%%%%%%%
Table \ref{tab:1} lists the authors and energy range of the reported absolute vertical intensity measurements. It is also shown if the experiment has been used by other reviews, namely {}``B''  \cite{buga}, {}``H\&T''  \cite{ht} and in the Particle Data Group {}``PDG''    \cite{rpp}. For H\&T we report also the final normalization factor they have found.
The symbol (*) refers to experiment using superconducting magnet spectrometers. 
%%%%%%%%%%%%%%%%%%%%%%%%%%%%%%%%%%%%%%%%%%%%%%%%%%%%%%%%%%
\begin{table}[t] 
\begin{center} 
\begin{tabular}{|l|c|c|c|c|} 
\hline
Reference & $p_\mu$ [GeV/c] & B & H\& T & PDG \\
\hline
\cite{aur} & 15.1 - 82.1 & X & 0.79 & \\
\cite{ayr} & 20-500 & X & & X \\
\cite{bab} & 11 - 810 & X & X & \\
\cite{ras} & 3- 3000 & X & 0.933 & X \\
\cite{bat} & 10-150 & X & 0.858 & \\
\cite{all} & 20-1000 & X & 1.039 & X \\
\cite{dep}* & 0.25-100 & X & 0.944 & X\\
\cite{kre}* & 0.2-120 &  & 0.818 & X\\
\cite{ach} & 20-3000 &  &  & X\\
\cite{hai}* & 0.6-400 &  &  & X\\
\hline
\end{tabular} 
\caption{\label{tab:1}
Compilation of different measurements of the muon momentum spectrum. Experiments with magnetic spectrometers are indicated with *. The second column reports the momentum range; the 3rd,4th and 5th columns if the data are used in the compilation of \cite{buga}, \cite{ht} (with the used normalizations factor) and \cite{rpp}, respectively.   } 
\end{center}
\end{table}
%%%%%%%%%%%%%%%%%%%%%%%%%%%%%%%%%%%%%%%%%%%%%%%%%

\begin{figure}[t]
\vspace*{2mm}
\begin{center}
\includegraphics[width=8.8cm]{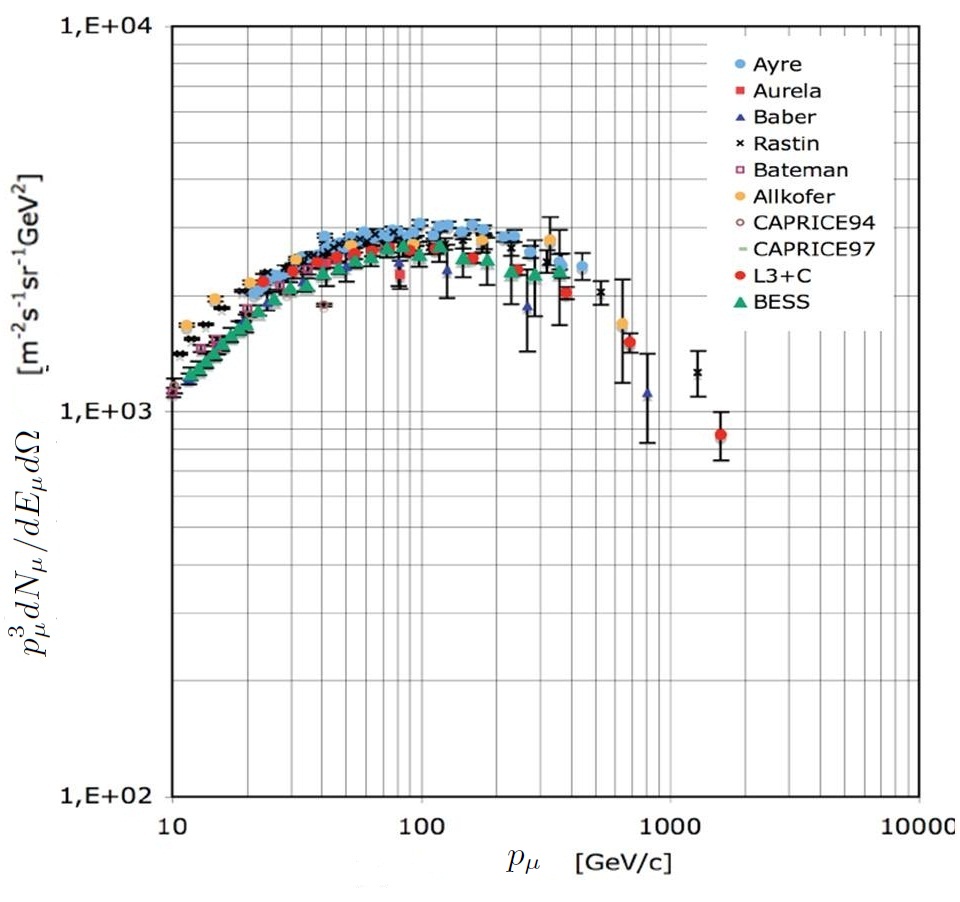}
\end{center}
\caption{\label{fig:pmu}
Vertical absolute differential muon intensity $ p_\mu^3 dN_\mu / dE_\mu d\Omega$ at sea level for the experiments reported in Tab. \ref{tab:1}. The ordinate values have been multiplied by $p_\mu^3$ in order to compress the plot and to emphasize the differences. In this energy range, $p_\mu \simeq E_\mu$.}
\end{figure}

The measurements listed in Table \ref{tab:1} (only published results) of the muon momentum from the vertical direction are presented in Fig. \ref{fig:pmu}. 
The agreement between measurements is relatively good and the largest contribution to the deviations are the systematic errors due to incorrect knowledge of the acceptance, efficiency of the counters and corrections for multiple scattering.
%%%%%%%%%%%%%%%%%%%%%%
%%%%%%%%%%%%%%%%%%%%%%%

\begin{figure}[t]
\vspace*{2mm}
\begin{center}
\includegraphics[width=10.0cm]{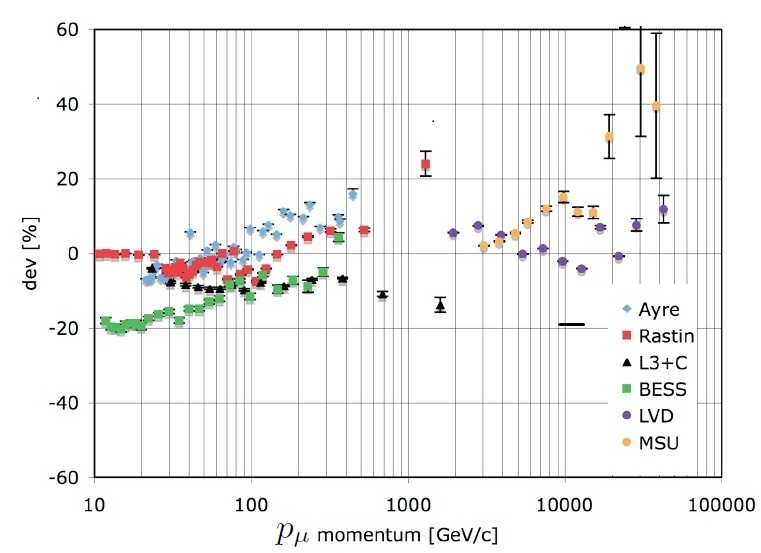}
\end{center}
\caption{\label{fig:devia}
Relative deviations of the differential muon intensity measurements at sea level with respect to the \cite{buga} parameterization. The symbols are the same as in Fig. \ref{fig:pmu}; the indirect measurements described in \S \ref{s:dir} and Fig. \ref{fig:pind} are also included.
}
\end{figure}

Measurements of the muon momentum spectra for $p_\mu < 1$ TeV/c are particularly important for the comparison of nuclear cascade models with available data.  
The differences between the measurements of the sea-level spectra are more evident if the percentage deviations of the data from one of the  parameterization are plotted (see Fig. \ref{fig:devia}).
Differences of $\pm$ 15-20\% exist between the data and the parameterization. The disagreement between the different experiments can be as large as 30-35\% despite the fact that the declared individual errors are small (however increasing with momentum due to the decreasing number of detectable particles and to the maximum detectable momentum).
The origin of the discrepancies can well be due to the incorrect knowledge/control of systematic effects. For example, in the case of L3+C the stable negative deviations can be due to a bad correction for the molasses that cover the apparatus.

%%%%%%%%%%%%%%%%%%%%%%
%%%%%%%%%%%%%%%%%%%%%%%

%%%%%%%%%%%%%%%%%%%%%
%%%%%%%%%%%%%%%%%%%%%
\section{Underground muons\label{sec:under}}
%%%%%%%%%%%%%%%%%%%%%
%%%%%%%%%%%%%%%%%%%%%

Underground measurements offer the possibility to extend the energy range of the muon spectrum beyond 1 TeV. Such data are of an indirect type, but their link with the direct lower energy observations gives the possibility to complete the picture of muon spectra measurements.

Deep underground detectors have normally large collecting area and are not subject to the time restrictions of balloon and satellite experiments, so they can measure the muon flux for a very long time. As a drawback, as discussed in \S 2, the energy dependence of the muon spectrum at $E_\mu >$ 1 TeV steepen one power and the intensity is a factor 1000 smaller than that of primary CRs on the upper atmosphere. 
The penetrating component of CRs underground depends on a complex convolution of different physics processes, as the high energy muon production spectrum and the muon energy losses. Particularly important is the knowledge of the composition and thickness of the material overburden the detector. 

%%%%%%%%%%%%%%%%%%%%%
\subsection{Muon energy losses}
%%%%%%%%%%%%%%%%%%%%%
Muon energy losses are usually divided into continuous and discrete processes. The former is due to ionization, which depends weakly on muon energy and can be considered nearly constant for relativistic particles. For muons below $\sim 500$ GeV, this is the dominant energy loss process. 
For muons reaching great depths, discrete energy losses become important: bremsstrahlung ($br$), direct electron-positron pair production ($pair$) and electromagnetic interaction with nuclei (photoproduction, $ph$). In these radiative processes energy is lost in bursts along the muon path. In general the total muon energy loss is parameterized as: 
\begin{equation}
\label{eq:muloss}
{dE_\mu \over dX} = -\alpha-\beta E_\mu
\end{equation}
where $X$ is the thickness of crossed material in g/cm$^2$ and $\beta = \beta_{br} + \beta_{pair} + \beta_{ph}$ is the sum of fractional energy loss in the three mentioned radiation processes. 
As the rock compositions are different for different underground experiments, the so-called \textit{standard rock} is defined as a common reference.
The standard rock is characterized by density $\rho= 2.65$ g/cm$^3$, atomic mass A = 22 and charge Z = 11. The thickness $X$ is commonly given in units of meters of water equivalent (1 m.w.e.= 10$^2$ g/cm$^2$).

The factors $\alpha$ and $\beta$ in Eq. (\ref{eq:muloss}) are mildly energy dependent as well as dependent upon the chemical composition of the medium: in particular $\alpha\propto Z/A$ and $\beta \propto Z^2/A$. Typical values are $\alpha \simeq 2$ MeV g$^{-1}$ cm$^{2}$ and $\beta \simeq 4\times 10^{-6}$ g$^{-1}$ cm$^{2}$. The critical energy is defined as the energy at which ionization energy loss equals radiative energy losses: $\epsilon_\mu=\alpha/\beta \simeq 500$ GeV.

\begin{figure}[t]
\vspace*{2mm}
\begin{center}
\includegraphics[width=10.0cm]{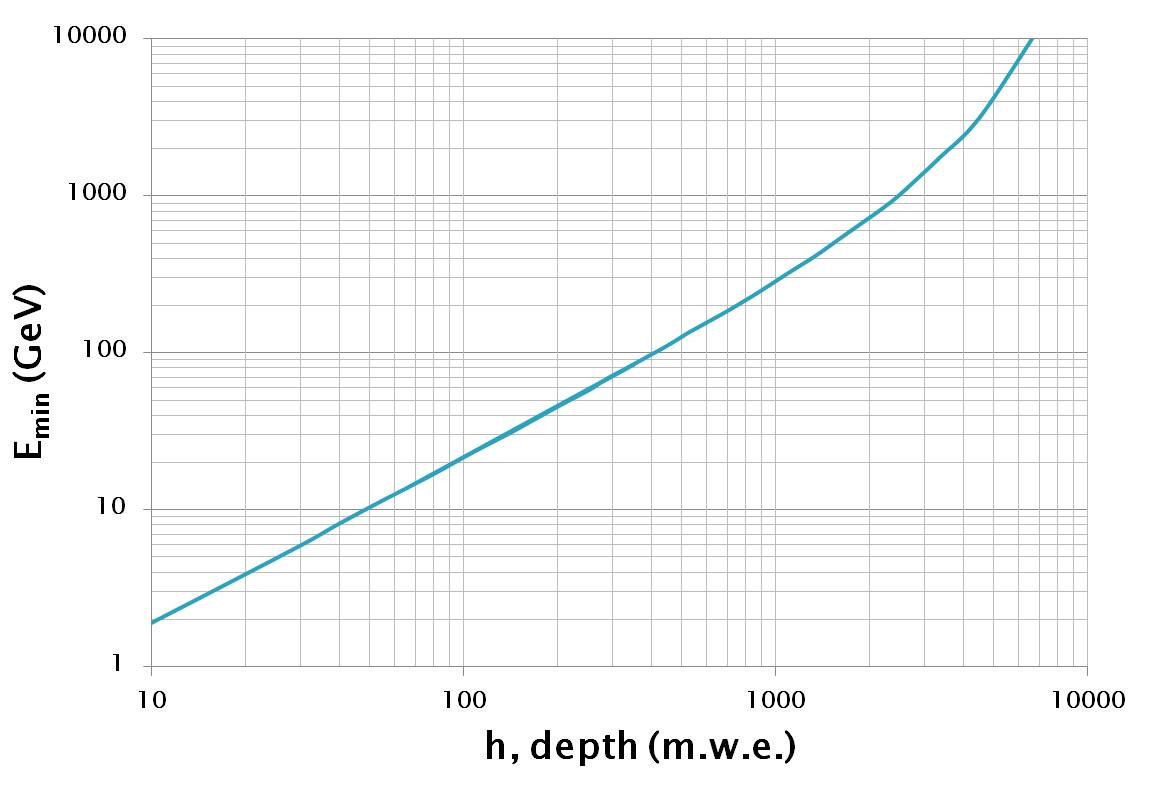}
\end{center}
\caption{\label{fig:e-vs-h}
Minimum energy at surface, $E_{min}$ to reach a given underground depth $X=h$. Typical values of $E_{min}$ for muons are 2, 20 and 2000 GeV to reach the ground (10 m.w.e.), the LEP tunnel ($\sim$80 m.w.e.) and the Gran Sasso Laboratory ($\sim$3 km.w.e.), respectively. }
\end{figure}

The general solution of Eq. (\ref{eq:muloss}) corresponds to the average energy $\langle E_\mu \rangle$ of a beam of muons with initial energy $E_\mu^0$ after penetrating a depth $X$:
\begin{equation}
\label{eq:mur1}
\langle E_\mu(X) \rangle = (E_\mu^0 + \epsilon_\mu) e^{-\beta X}-\epsilon_\mu
\end{equation}
The minimum energy required for a muon at the surface to reach slant depth $X$ is the solution of Eq. (\ref{eq:mur1}) with residual energy $E_\mu(X)=0$:
\begin{equation}
\label{eq:mur2}
E^0_{\mu, min}  = \epsilon_\mu (e^{\beta X}-1) 
\end{equation}
The range $R$ for a muon of energy $E_\mu^0$, i.e. the underground depth that this muon will reach, is:
\begin{equation}
\label{eq:mur3}
R(E^0_\mu)  ={1\over \beta} \ln(1+{E_\mu^0\over \epsilon_\mu}) 
\end{equation}
The above quantities are average values. For precise calculations of the 
underground muon flux one needs to take into account the fluctuations inherent to the radiative processes. Because of the stochastic character of muon interaction processes with large energy transfers (e.g., bremsstrahlung) muons are subject to a considerable range straggling. The
higher $E^0_\mu$ is, the more dominant are the radiation processes and the more important are the fluctuations of the energy losses which broaden the distribution of the range. Fig. \ref{fig:e-vs-h} shows the typical values of the minimum energy at surface, $E_{min}$ to reach a given underground depth.
%%%%%%%%%%%%%%%%%%%%%%
%%%%%%%%%%%%%%%%%%%%%%%

%%%%%%%%%%%%%%%%%%%%%
\subsection{The deep-intensity relation\label{s:dir}}
%%%%%%%%%%%%%%%%%%%%%

The muon spectrum at energies above few TeV was deduced by underground measurements.  The procedure used for this \textit{indirect} measurement of the sea-level energy spectrum passes through the determination of the so-called \textit{deep-intensity relation} (DIR) $I^0_\mu(h)$. This quantity represents the muon intensity at the vertical direction as a function of the depth $h$. As underground detectors are at a fixed position, in principle only one point can be measured. However, when measuring the muon intensity $I_\mu(h,\theta)$ at different zenith angle $\theta$, the quantity of rock (or water) overburden changes. 
At high energy ($E_\mu> 1$ TeV) and for $\theta<60^\circ$, Eq. (\ref{eq:gaisser}) provides a simple relationship between $I^0_\mu(h)$ and $I_\mu(h,\theta)$:
\begin{equation}
\label{eq:Ih}
I_\mu(h,\theta)= I^0_\mu(h)/\cos\theta
\end{equation}
From the experimental point of view, $I_\mu(h,\theta)$ is measured at a given direction $\theta$, corresponding to a slant depth $h$, as:
\begin{equation}
\label{eq:Iht}
I_\mu(h,\theta)= \biggl( {1\over \Delta T} \biggr) {\sum_i N_i m_i \over \sum_j \Delta \Omega_j(\theta,\phi) A_j(\theta,\phi) \epsilon_j(\theta,\phi) }
\end{equation}
where $\Delta T $ in the total livetime (in s) of the experiment, $N_i$ is the number of detected events with multiplicity $m_i$ with zenith $\theta$ and azimuth $\phi$ in the angular bin $\Delta \Omega_j(\theta,\phi)$ (sr).
 $A_j(\theta,\phi)$ (m$^2$) and $\epsilon_j(\theta,\phi)$ correspond to the geometrical intrinsic acceptance of the detector, and the overall efficiency in the $(\theta,\phi)$ bin. 

The relation between the measured DIR function $I^0_\mu(h)$ (measured from Eq. (\ref{eq:Iht}) with the use Eq. (\ref{eq:Ih})) and the differential sea-level muon spectrum ${dN_\mu \over dE_\mu d\Omega}$ is expressed as:
\begin{equation}
\label{eq:ImIh}
I^0_\mu(h)= \int^\infty_0 {dN_\mu \over dE_\mu d\Omega} P(E_\mu,h) dE_\mu
\end{equation}
Here, $ P(E_\mu,h) $ is the muon survival function. It represents the probability that muons of energy $E_\mu$ at surface reach a given depth $h$, and it is determined via Monte Carlo calculations. Assuming Eq. (\ref{eq:gaisser}) for the sea-level muon spectrum, leaving as free parameters the muon spectral index $\gamma$, the scale $A$ and the balance $B$ factors, it is possible to unfold the sea level muon spectrum from the measured vertical muon intensity. 
%%%%%%%%%%%%%%%%%%%%%%%%%%%%%%%%%%%%%%%%%%%%%%%%%%%%%%%%%%
\begin{table}[t] 
\begin{center} 
\begin{tabular}{|l|c|c|} 
\hline
Experiment  & Depth      & Momentum \\
(Reference) & ($m.w.e.$) & (TeV/c) \\
\hline
LVD \cite{lvd} &  $>$3000 & 1.9-43 \\
MACRO \cite{mac-dir} & $>$3150 & 0.5-20 \\
Baksan \cite{baksan} &  $>$850  & 1-30 \\
MSU \cite{msu} &  $>$50  & 3-50 \\
\hline
\end{tabular} 
\caption{\label{tab:2}
Compilation of different indirect measurements of the muon momentum with underground detectors.} 
\end{center}
\end{table}
%%%%%%%%%%%%%%%%%%%%%%%%%%%%%%%%%%%%%%%%%%%%%%%%%

\begin{figure}[t]
\vspace*{2mm}
\begin{center}
\includegraphics[width=10.0cm]{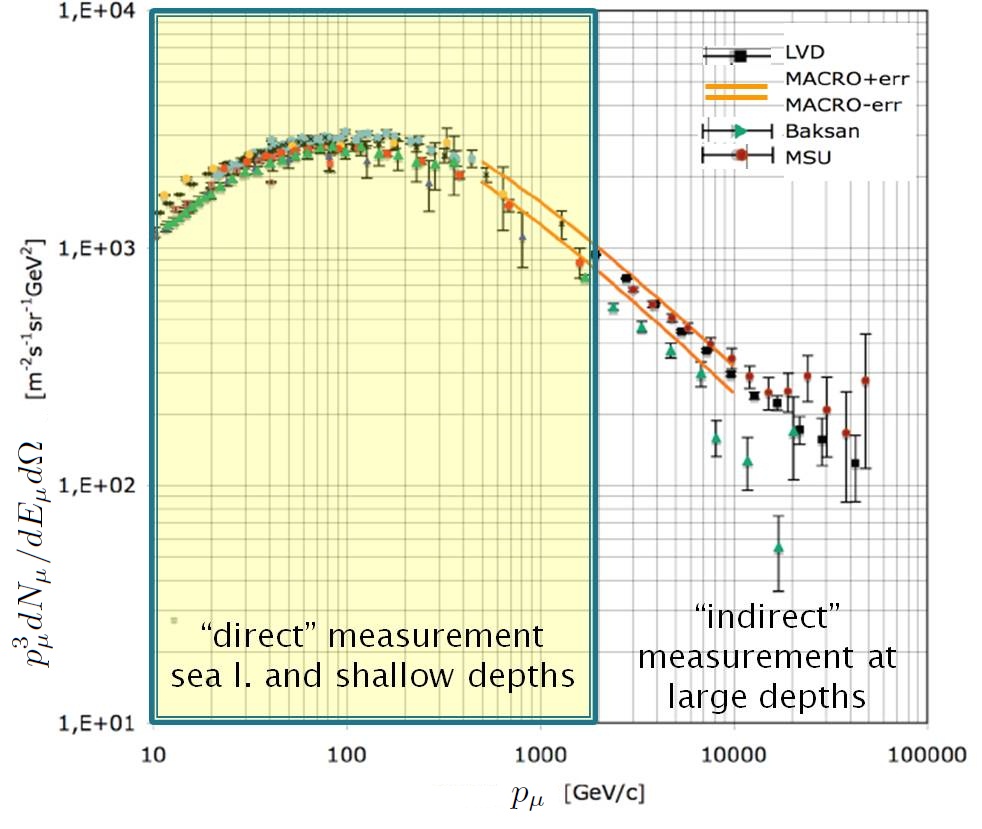}
\end{center}
\caption{\label{fig:pind}
 Extension of the data shown in Fig. \ref{fig:pmu} to higher momenta, by including the indirect measurement of the underground muon flux.}
\end{figure}

Table \ref{tab:2} indicates the experiments, the depth and the estimated muon momentum range for these indirect underground measurements. The results are shown in Fig. \ref{fig:pind}. 
The differential distribution ($dN_\mu/dE_\mu d\Omega$) is presented, as usual, multiplied by a factor $p_\mu^3$ (as momentum and energy coincide) to better observe the variation of the spectrum in the whole energy region. In these indirect measurements, the main sources of systematic uncertainties are due to the treatment of hard processes in the energy loss of muons and to the knowledge of the rock density and overburden thickness, which rely on geological surveys. 
%%%%%%%%%%%%%%%%%%%%%%
%%%%%%%%%%%%%%%%%%%%%%%

%UNDERWATER 
This situation seems to favour underwater/ice experiments. In this case, the uncertainty in the material density and overburden is negligible with respect to underground experiments. However, when small-size experiments are carried out, the main uncertainty is the multiplicity of the muon bundle. In the case of large underwater experiments (like the running neutrino telescopes, \cite{chia}), they are optimized to look for upward-going neutrino induced particles. Atmospheric muons are seen with the {}``tail of the eyes'' (the photomultipliers inside optical modules), where large uncertainties on the optical module angular acceptance do not allow a precision measurements. See \cite{anta-5} for a discussion and the DIR measurement with the ANTARES underwater neutrino telescope. 

%%%%%%%%%%%%%%%%%%%%%%%%%%%%%%%%%%%%%%%%%%%%%%%%%%%%%%%%%%%%%%%%%%%%%
\section{Characteristic of underground/underwater muons\label{sec:undermu}}
%%%%%%%%%%%%%%%%%%%%%%%%%%%%%%%%%%%%%%%%%%%%%%%%%%%%%%%%%%%%%%%%%%%%%
%%%%%%%%%%%%%%%%%%%%%%
\begin{figure}[tbh]
\vspace*{2mm}
\begin{center}
\includegraphics[width=14cm]{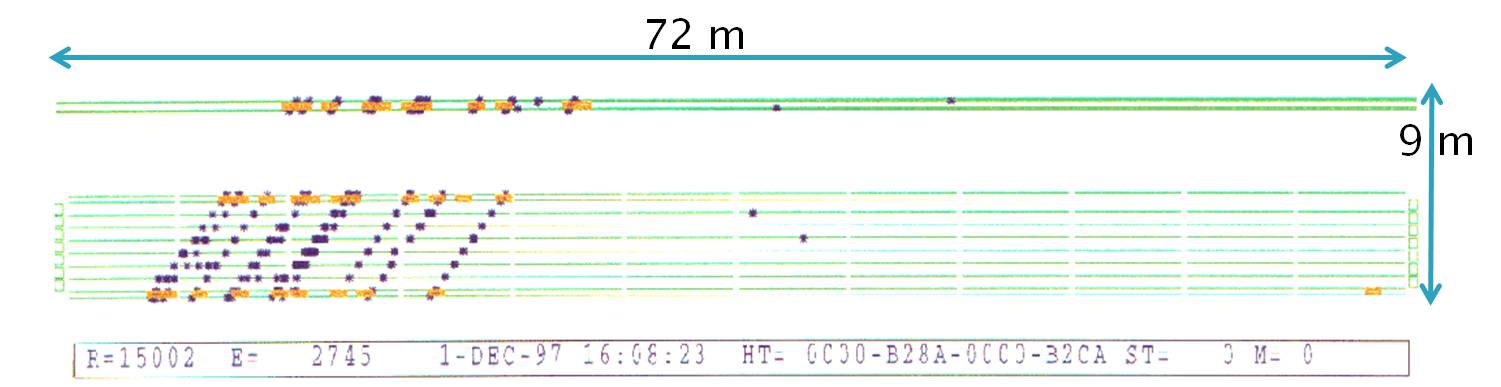}
\end{center}
\caption{\label{fig:multimu}
 A multimuon event seen in the MACRO experiment. 10 different tracks are identified.}
\end{figure}
%%%%%%%%%%%%%%%%%%%%%%%
Fig. \ref{fig:multimu} shows a multiple muon event detected by MACRO. 
Multiple events are closely packed bundles of parallel muons, usually of high energy, originating from the same primary CR. Multi-muon events are used to explore the properties of very high energy hadronic interactions and to study the longitudinal development of showers \cite{mac-mu}. 
The multiplicity of produced secondary particles increases with the energy of the initiating particle.
The muon multiplicity is an observable which is also correlated with the mass of the primary CR: at a given total energy, heavier nuclei produce more muons than a primary proton.

The interaction vertex of the particles which initiate the air showers is typically at an atmospheric altitude of 15 km. Since secondary particles in hadronic cascades have small transverse momenta $p_t$ ($\sim$ 300 MeV/c), high energy muons are essentially collimated near the shower axis. Considering a primary nucleon, producing mesons of energy $E_{\pi,K}$ with transverse momentum $p_t$ at a height $H_{prod}$, the average separation of their daughter high energy muons from the shower axis is given by \cite{mac-mupair}:
\begin{equation}
\label{eq:r-pt}
r\simeq {p_t \over E_{\pi,K} } H_{prod}
\end{equation}
For primary energies around 10$^{14}$ eV, the lateral displacements of energetic muons ($\sim 1$ TeV) of several meters are typically obtained underground. Displacements are almost exclusively caused by transferred transverse momentum in hadronic processes. Typical multiple scattering angles for muon energies around 100 GeV in thick layers of rock (50-100 m) are on the order of a few mrad.

%%%%%%%%%%%%%%%%%%%%%
\subsection{Atmospheric muons in neutrino telescopes \label{sec:watermu}}
%%%%%%%%%%%%%%%%%%%%%

Atmospheric muons represent the most abundant signal in a neutrino telescope and can be used to calibrate the detector and to check its expected response to the passage of charged particles. On the other side, they can represent a dangerous background source because downward-going muons can incorrectly be reconstructed as upward-going particles and mimic high energy neutrino interactions; muons in bundles are particularly dangerous. These muons are expected to arrive almost at the same time in the plane perpendicular to the shower axis. A full Monte Carlo simulation, starting from the simulation of atmospheric showers, can accurately reproduce the main features of muons reaching a neutrino telescope, but requires a large amount of CPU time. 

Recently, parametric formulae \cite{para} to evaluate the flux of atmospheric muons were derived from a full Monte Carlo simulation. These formulas take into account the muon multiplicity and the energy spectrum of muons in a bundle, as a function of the distance from the shower axis. A simple generator interface is provided (called MUPAGE), which can be used by all experiments having a flat overburden coverage of at least 1500 m.w.e. \cite{mupage}. MUPAGE is used to simulate atmospheric muons in  Mediterranean neutrino telescopes. 

\begin{figure}[tb]
\vspace*{2mm}
\begin{center}
\includegraphics[width=9.8cm]{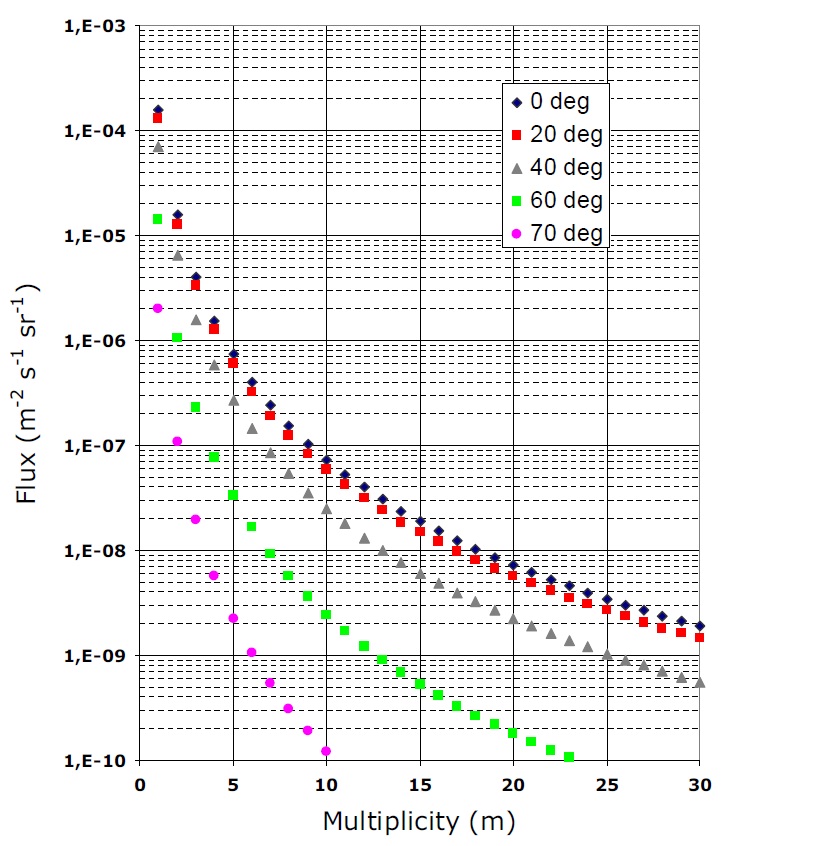}
\end{center}
\caption{\label{mupage_mul}
Flux of bundles of muons as a function of their multiplicity $m$ in the bundle obtained with Eq. (\ref{eq:eq1}) at the depth of 3 km.w.e. and for 5 different values of the zenith angle. The computation assumes here a muon energy threshold of 1 GeV.}
\end{figure}

The flux (which corresponds to the $(dN_\mu / dE_\mu d\Omega)$ of Eq.  (\ref{eq:gaisser}) integrated over  the muon energy $E_\mu$, with units: $m^{-2} s^{-1} sr^{-1}$) of muon bundles with multiplicity $m$ (see Fig. \ref{mupage_mul}) is obtained as a function of the depth along the vertical direction $h_0$ (note that the index 0 means that the depth $h$ is computed exactly at the vertical direction) and zenith angle $\theta$ as:
\begin{equation}
\Phi(m;h_0,\theta)= {K(h_0,\theta) \over m^{\nu(h_0,\theta)}} 
\label{eq:eq1}
\end{equation}
The flux of bundles of increasing multiplicity $m$ decreases with increasing vertical depth and zenith angle. The parametric dependences of $ K(h_0,\theta)$ and $\nu(h_0,\theta)$ are reported in \cite{para}, as the others below.
%%%%%%%%%%%%%%%%%%%%%%
%%%%%%%%%%%%%%%%%%%%%%%

The energy spectrum of muons is described \cite{gai90} by:
\begin{equation}
{ dN \over d (log_{10}E_\mu) } = G E_\mu e^{\beta X (1-\gamma)} [E_\mu + \epsilon (1-e^{-\beta X})]^{- \gamma}    
\label{eq:spectrum} 
\end{equation}
where $X=h_0/\cos\theta$, $\gamma$ is the spectral index of the primary CRs and $\epsilon= \alpha /\beta$; $\alpha ,\beta$ are defined in Eq. (\ref{eq:muloss}). In the parameterization, $\gamma$ and $\epsilon$ were instead considered as free fit parameters. 
The constant $G=G(\gamma,\epsilon)$ represents a normalization factor, in order that the integral over the muon energy spectrum (\ref{eq:spectrum})
from 1 GeV  to 500 TeV is equal to 1. In the case on single muon events (bundles with $m=1$) a simple dependence $\gamma=\gamma(h_0)$ and $\epsilon=\epsilon(h_0,\theta)$ holds. 

The situation is more complicated for multiple muons. Due to the muon production kinematics, the muon energy depends from their distance with respect to the axis of the bundle. The description of the muon lateral distance $R$ from the axis is thus the preliminary step to evaluate the muon energy distribution in a bundle. $R$ (in the plane orthogonal to the axis) was extracted from a distribution of the form: 
\begin{equation}
{dN \over dR} = C { R\over (R+R_0)^\alpha }
\label{eq:radial} 
\end{equation}
where $R_0=R_0(h_0,m,\theta)$ and $\alpha=\alpha(h_0,m)$.
The energy spectrum of muons arriving in bundles has the same general form as for single muons (\ref{eq:spectrum}). In the case of multiple muons, the analytic description of the parameters $\gamma=\gamma(h,R,m)$ and $\epsilon=\epsilon(h,R,\theta)$ depends on 15 constants.

\begin{figure}[tb]
\vspace*{2mm}
\begin{center}
\includegraphics[width=8.6cm]{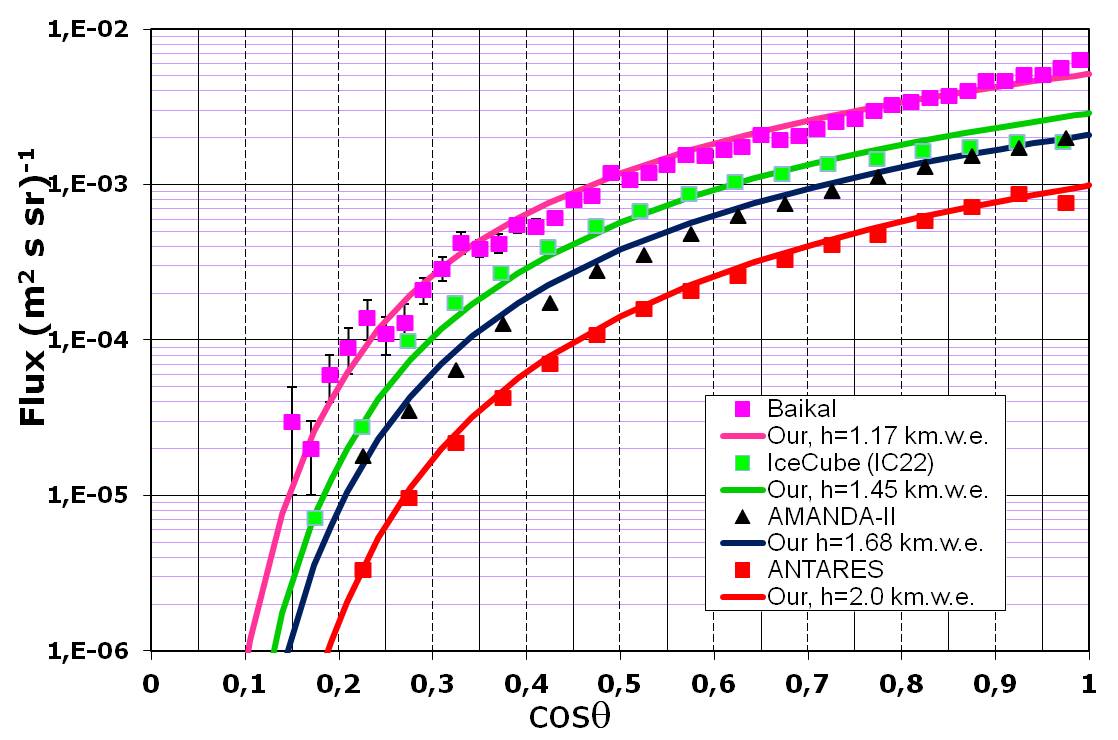}
\end{center}
\caption{\label{mu-cosze}
Muon flux as a function of the $\cos\theta$ as measured by ANTARES\cite{anta-5}, AMANDA-II \cite{amanda}, IceCube \cite{icecube}, and Baikal \cite{baikal} at four different depths. The results of the MUPAGE computation is superimposed as full lines.
}
\end{figure}

These parameterizations allow to evaluate not only the total muon flux, but also the total number of muon bundles in deep detectors starting from the primary CR flux, CR composition and interaction
model which reproduces (at the level of $\sim$ 30\%) the MACRO data (depth: 3000 -6000 m.w.e., $\theta < 60^\circ$). Fig. \ref{mu-cosze}
 shows the comparisons of the zenith distribution evaluated at a fixed depth using Eq. (\ref{eq:eq1}) and some underwater/ice data.

%%%%%%%%%%%%%%%%%%%%%%
%%%%%%%%%%%%%%%%%%%%%%%

%%%%%%%%%%%%%%%%%%%%%
\subsection{Pressure and temperature effect\label{sec:tp}}
%%%%%%%%%%%%%%%%%%%%%

If atmospheric temperature changes by $\Delta T$, the muon flux at the observation level $X$ will change by a quantity $\Delta I_\mu$. This variation can be either positive or negative, i.e. $\Delta I_\mu\propto \pm \Delta T$, depending on the energy $E_\mu$ of the observed muon. Two competitive effects are in action if the atmospheric temperature increases. As a consequence, the atmosphere expands and the air density  decreases; the probability of the interaction of mesons (kaons and pions) at unit of geometric path becomes smaller, hence decay probability into muons becomes higher. On the other hand, the geometric expansion of the atmosphere increases the path from the generation point to the detector, and a higher number of muons will decay.
For low energy muons, the latter effect is the dominant one, and the correlation sign between flux and temperature is negative ($\Delta I_\mu\propto - \Delta T$). For high energies, muons have not enough time to decay in the atmosphere and the correlation sign becomes positive \cite{dmitri}.

\begin{figure}[tb]
\vspace*{2mm}
\begin{center}
\includegraphics[width=10.5cm]{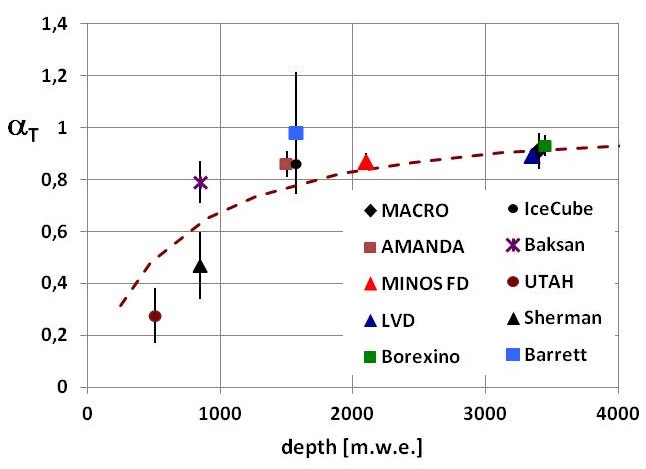}
\end{center}
\caption{\label{fig:at}
The temperature coefficient $\alpha_T$ as a function of detector depth. The dashed curve is the prediction using the pion-only model \cite{mac-t}. The points correspond (from top to bottom): \cite{mac-t},\cite{buchta}, \cite{gras}, \cite{selvi}, \cite{bellini}, \cite{desiati}, \cite{andre}, \cite{cut}, \cite{she}, \cite{barrett}.   }
\end{figure}

Many under-ground/water/ice experiments measured the correlation between temperature and high energy muon intensity. It was found that \cite{mac-t} an effective temperature $T_{eff}$, defined by the weighted average of temperatures from the surface to the top of the atmosphere, is useful to describe the situation. $T_{eff}$ approximates the atmosphere as an isothermal body, weighting each pressure layer according to its relevance to muon production in atmosphere. The variation of muon rate $\Delta I_\mu/I_\mu$ is related to the effective temperature as:
\begin{equation}
{\Delta I_\mu \over I_\mu} = \alpha_T { \Delta T_{eff}\over T_{eff} }
\label{eq:T1} 
\end{equation}
where $\alpha_T$ is the atmospheric temperature coefficient, which is a function of both the muon threshold energy and the $K/\pi$ ratio. As the energy increases, the muon intensity becomes more dependent on the
meson critical energy $\epsilon_{\pi,K}$, which in turn depends on the atmospheric temperature. The $\alpha_T$ coefficient reflects the fraction of mesons that are sensitive to atmospheric temperature variations. For energies much greater than the critical energy, and thus for very deep experiments, the value of $\alpha_T$ approaches unity.
The expected effective temperature coefficient as a function of depth is shown in Fig. \ref{fig:at}, together with the values measured by underground/ice experiments.
%%%%%%%%%%%%%%%%%%%%%%
%%%%%%%%%%%%%%%%%%%%%%%

%%%%%%%%%%%%%%%%%%%%%
\section{Muons and neutrinos\label{sec:nuEmu}}
\begin{figure}[t]
\vspace*{2mm}
\begin{center}
\includegraphics[width=10.8cm]{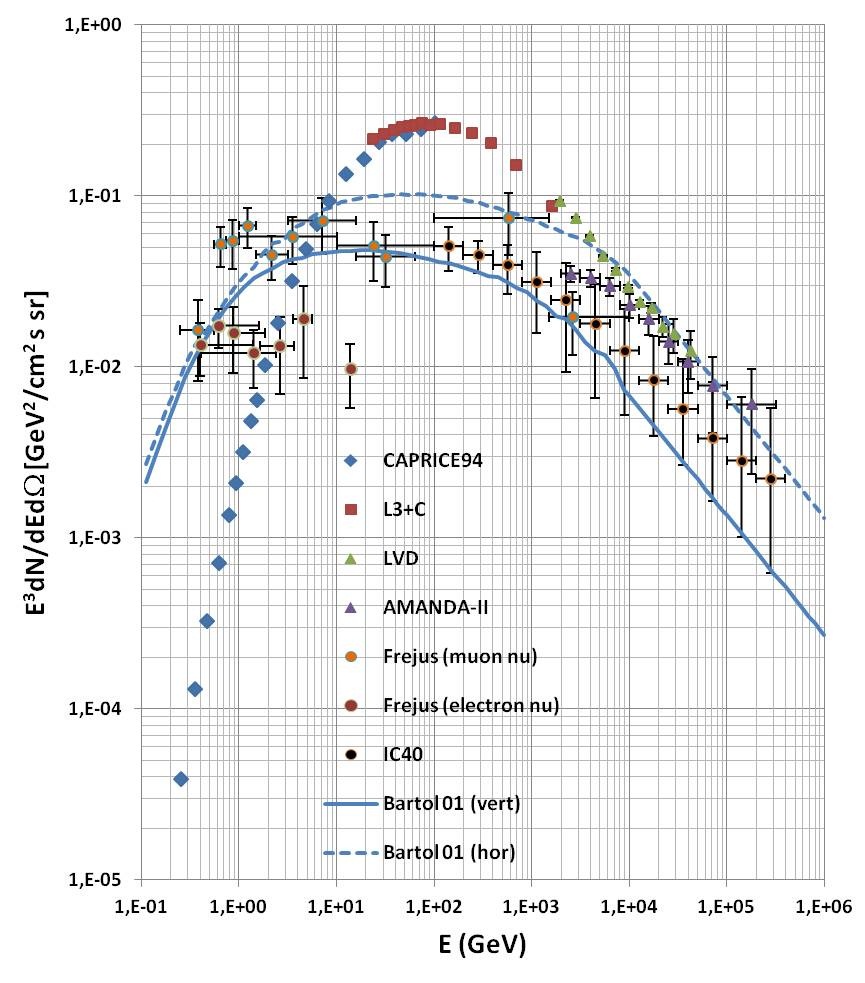}
\end{center}
\caption{\label{fig:nu-mu}
Compilation of measurements of atmospheric muons and neutrinos. The points correspond (from top to bottom): \cite{dep}, \cite{ach}, \cite{lvd}, \cite{amanda2}, \cite{daum}, \cite{ic40}. The lines represent the neutrino flux expectation (full: vertical direction; dashed: horizontal direction) computed by \cite{barr}.
 }
\end{figure}

%%%%%%%%%%%%%%%%%%%%%
As indicated in Eq. (\ref{eq:meson}), the production mechanisms of atmospheric muons and neutrinos are strongly correlated, see \cite{illana} for a recent review. However, due to the two-body kinematics, the energy spectra of the $\mu $'s and $\nu_\mu$'s from mesons decay are different. 
Let us consider for instance the pion decay in the centre of mass (c.m.) system ($m_\pi=139.6$ GeV; $m_\mu=105.7$ GeV). The c.m. muon energy is $E_\mu^*= {(m_\pi^2+m_\mu^2)/ 2m_\pi} =109.8\textrm{ MeV}$.
Similarly for the neutrino, considering that in the c.m. system $E_\mu^*+E_\nu^*=m_\pi$, one has:
$E_\nu^*= (m_\pi^2-m_\mu^2)/ 2m_\pi = 29.8 \textrm{ MeV}$.
In the laboratory system, the energies are boosted by the Lorentz factor $\gamma = E_\pi/m_\pi$. In any case, muons carry a larger fraction of the meson energy than neutrinos. As consequence, the energy distribution of $\nu_\mu$ is slightly shifted towards lower energy values than charged muons, as shown in Fig. \ref{fig:nu-mu}. 
Additional $\nu_\mu$ are produced by the in flight decay of muons, together with a $\nu_e$ and an electron/positron. As the muon decay probability in the atmosphere decreases with increasing $E_\mu$, the $\nu_e$ spectrum is depleted with respect to that of $\nu_\mu$ at high energy. 
In Fig. \ref{fig:nu-mu}, we include the measurement of the $\nu_\mu$ energy spectrum reported by the Frejus, AMANDA and IceCube experiments.
The $\nu_e$ component was measured between $\langle E_\mu \rangle \sim$0.4 - 14 GeV by the Frejus experiment.  
The measurement of the muon and neutrino energy spectra represents a very challenging result, as completely different experimental techniques were used to measure the charged and neutral leptons in different energy ranges.

%% ILLANA
Neutrino telescopes are taking data under the Antarctic ice or in the Mediterranean sea. The main goals are the detection of neutrinos from cosmic sources and the measurement of the isotropic flux of high energy neutrinos from the ensemble of all extragalactic sources. The signature for the former is an excess of events over the background of atmospheric neutrinos from a given direction. The signatures of the diffuse astrophysical neutrino signal are: (a) isotropy; (b) a hard energy spectrum; (c) approximately equal fluxes of $\nu_e, \nu_\mu$ and $\nu_\tau$. The neutrino fluxes generated by the \textit{prompt} charm decay have also the properties (a) and (b), and equal fluxes for $\nu_e$ and $ \nu_\mu$ and therefore constitute a dangerous background. 
%Most theoretical models for the production of astrophysical neutrinos %predict an energy spectrum harder than what is expected for the charm %decay component; however these predictions have important uncertainties.

%%%%%%%%%%%%%%%%%%%%%%
%%%%%%%%%%%%%%%%%%%%%%%

%%\conclusions  
\section{Conclusions}
%% \conclusions[modified heading if necessary]
The energy spectrum of atmospheric muons is measured over almost 6 decades of muon momentum with different experimental techniques at sea level and in underground/water/ice experiments. The so-called \textit{prompt} component, which is expected to dominate the spectrum at very high energies, is still undetected. 
Below 1 TeV/c, disagreements between different experiments are up to 20\% due to systematic uncertainties. Slightly larger uncertainties arise from the indirect methods used to deconvolve the higher energy spectrum up to $\sim 40$ TeV. These measurements of the energy spectrum above few TeV were made with large underground detectors. In particular, the MACRO experiment at the Gran Sasso Laboratory in Italy accurately studied 
the multiplicity distribution of muons arriving in bundle, their lateral distribution with respect to the shower axis and their energy spectrum. These information were used to obtain a parameterization of the atmospheric muons detected by large underwater neutrino telescopes. 

The energy spectrum of atmospheric muon neutrinos are measured through the detection of upgoing muons, generated by charged current interactions of atmospheric $\nu_\mu$. They represent the irreducible background for searches of cosmic neutrinos, and for this reason this component must be accurately known. The $\nu_e$ component has still to be accurately measured, particularly in the high energy region.

The knowledge of the atmospheric muon spectrum, the characteristics of the muon flux at sea level and the processes of attenuation of muons passing through rocks or other materials is of fundamental importance for Earth science purposes and muon imaging feasibility.

%%\begin{acknowledgements}
\vskip 0.6cm
\noindent\textbf{Acknowledgements}
M.S. thanks the conference organizers for the invitation at the \textit{Muon and Neutrino Radiography 2012} (MNR 2012) hold in Clermont Ferrand. The authors would acknowledge the collaboration of the Bologna colleagues of the (former) MACRO experiment, and of the Opera and ANTARES collaborations. 
%\end{acknowledgements}

%%%%%%%%%%%%%%%%%%%%%%%%%%%%%%%%%%%%%%%%%%%%%%%%%%%%%%%%%%%%%%%%%%%%%%%%%
%%%%%%%%%%%%%%%%%%%%%%%%%%%%%%%%%%%%%%%%%%%%%%%%%%%%%%%%%%%%%%%%%%%%%%%%%
%%%%%%%%%%%%%%%%%%%%%%%%%%%%%%%%%%%%%%%%%%%%%%%%%%%%%%%%%%%%%%%%%%%%%%%%%
\end{document}